\documentclass[journal,twoside]{IEEEtran}

\usepackage{setspace}
\usepackage{multirow}
\usepackage{lipsum}
\usepackage{amsfonts}
\usepackage{array}
\usepackage{amsmath,cases,amssymb}
\usepackage{amsmath}
\usepackage{amsthm}
\usepackage{graphicx}
\usepackage{cite}
\usepackage{subfigure}
\usepackage{epsfig}
\usepackage[hyphens]{url}
\usepackage{algorithm}
\usepackage{algorithmic}
\usepackage{epstopdf}
\usepackage{color}
\usepackage{makecell}
\usepackage{xcolor}
\usepackage[normalem]{ulem}

\newcommand{\PA}{\mathtt{PA}}
\newcommand{\SA}{\mathtt{SA}}

\DeclareGraphicsExtensions{.eps,.pdf}

\usepackage{soul}

\makeatletter
\newcommand{\doublewidetilde}[1]{{%
		\mathpalette\double@widetilde{#1}}}
\newcommand{\double@widetilde}[2]{%
	\sbox\z@{$\m@th#1\widetilde{#2}$}%
	\ht\z@=.5\ht\z@
	\widetilde{\box\z@}}
\makeatother

\newcommand{\phuc}[1]{\textcolor{black}{#1}}

\begin{document}

\title{The Role of Game Networking in the Fusion of Physical and Digital Worlds through \\6G Wireless Networks}
	
\author{Van-Phuc Bui, \textit{IEEE Student Member}, Shashi Raj Pandey, \textit{IEEE Member}, Andreas Casparsen, \\ Federico Chiariotti, \textit{IEEE Member},  Petar Popovski, \textit{IEEE Fellow}  
		\thanks{Van-Phuc Bui (vpb@es.aau.dk), Shashi Raj Pandey (srp@es.aau.dk), Andreas Casparsen (aca@es.aau.dk), and Petar Popovski (petarp@es.aau.dk) are with the Department of Electronic Systems, Aalborg University, Denmark. Federico Chiariotti (chiariot@dei.unipd.it) is with the Department of Information Engineering, University of Padova, Italy. This work was supported by the Velux Foundation through the Villum Investigator grant ``WATER.'' Federico Chiariotti's work is supported by the European Union, under the Italian NRRP Young Researchers grant ``REDIAL.''}
		
	}
	\maketitle
\begin{abstract}
The sixth generation (6G) of wireless technology is seen as one of the enablers of real-time fusion of the physical and digital realms, as in Digital Twin,  eXtended reality, or the Metaverse. This would allow people to interact, work, and entertain themselves in an immersive social network of online 3D~virtual environments. From the viewpoint of communication and networking, this will represent an evolution of the \emph{game networking} technology, designed to interconnect massive users in real-time online gaming environments. This article presents the basic principles of game networking and discusses their evolution towards meeting the requirements of the Metaverse and similar applications. Several open research challenges are {discussed}, along with possible solutions through experimental case studies.
\end{abstract}
% \vspace{-0.5cm}

%%%%%%%%%%%%%%%%%%%%%%%%%%%%%%%%%%%%%%%%%%%%%%%%
\section{Introduction}\label{sec:intro}
%%%%%%%%%%%%%%%%%%%%%%%%%%%%%%%%%%%%%%%%%%%%%%%%

Digital Twins (DTs)  along with the Metaverse and eXtended Reality (XR), represents a trend toward a real-time fusion of the physical and the digital realms. The vision of the these concepts is to move activities such as play, work, and socializing within a virtual world~\cite{ning2021survey}, and more advanced industrial and consumer application aim to further integrate virtual elements and physical activities. The virtual world contains digital representations of individuals and objects in the physical world, facilitating social interaction and engagement in various contexts. Meeting the challenging connectivity requirements for this digital-physical fusion is seen as one of the main objectives for sixth-generation (6G) wireless technology, targeting services related to XR, VR/AR, haptics, and brain-computer interaction.

\begin{figure}[tp]
		\centering
		\includegraphics[trim=0cm 0cm 0cm 0cm, clip=true,width=1\columnwidth]{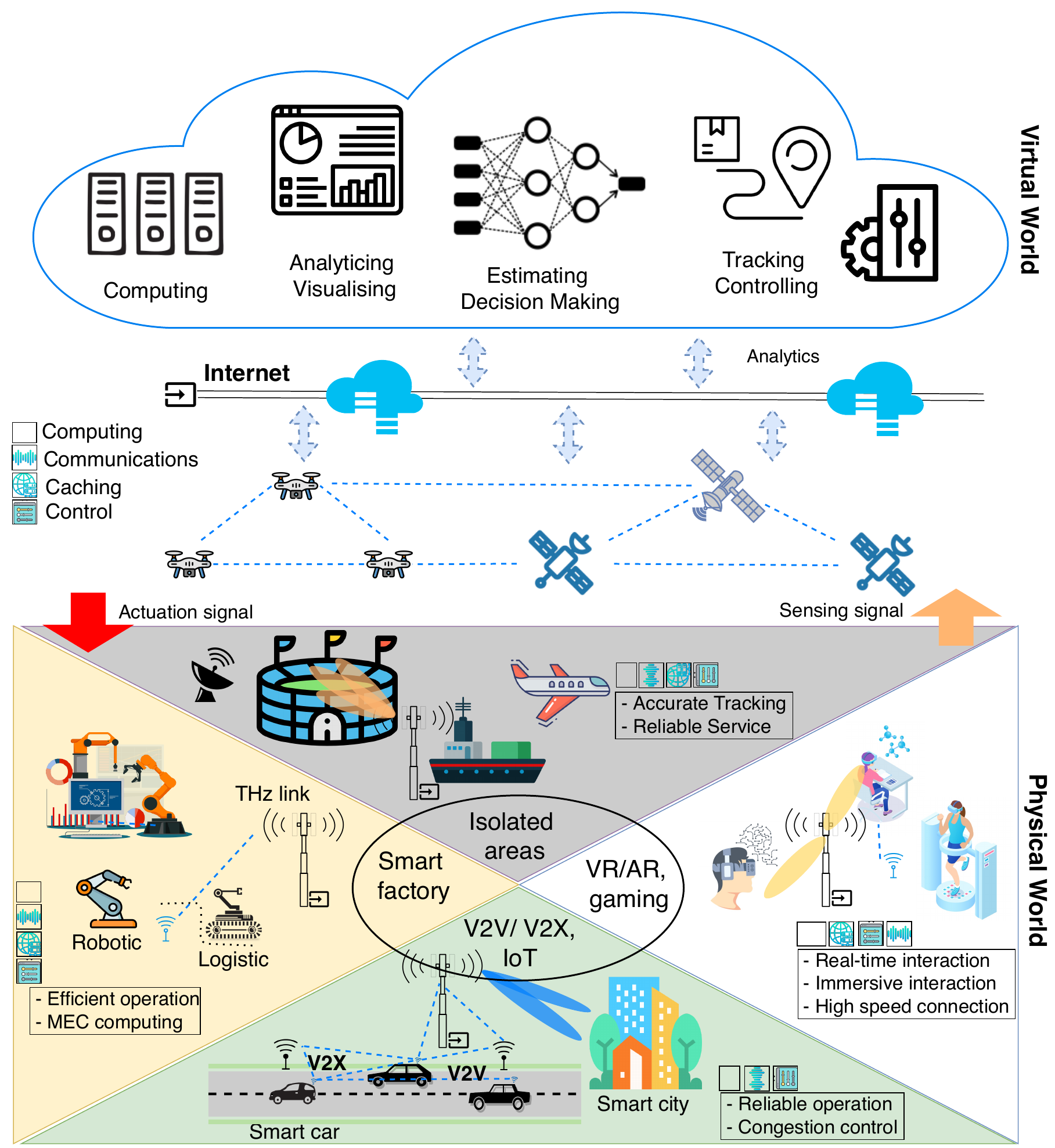}
   % \vspace*{-0.1cm}
		\caption{An illustration of networking infrastructure.}
		\label{network}
   % \vspace*{-0.5cm}
\end{figure}
Game Networking (GN) (also referred to as multiplayer networking or netcode) offers the proper point of departure to drive the evolution of the communication/networking technology towards supporting immersive social network and Metaverse requirements~\cite{ball2022metaverse}. It entails protocols and algorithms for sending and optimizing network packets as well as creating centralized authority to enhance user interactions, and may be considered a precursor to the communication/networking technology necessary for the Metaverse or similar concepts that offer a real-time integration of the physical and digital worlds~\cite{smed2017algorithms}. Recently, technology companies have been exploring various approaches to developing and implementing shared virtual and mixed spaces, leveraging massive game networking platforms as their foundation. 
Originating from one of the categories of computer networking that aims to provide the best end-user experience in throughput-intensive, latency-sensitive activities, GN has been developed to meet the demanding requirements of massive multi-user online games that deliver a real-time experience to the users. These stringent timing requirements go along with the requirements of 6G, calling for the incorporation of GN principles. As new technologies such as DT and VR/AR continue to emerge, future communications will evolve to encompass the physical world and the virtual world, bringing significant challenges in terms of synchronization between physical and virtual events, as well as between different players located all around the world. Maintaining the illusion of locality and a coherent timeline across thousands or millions of endpoints will stress even 6G's predicted capabilities. The features of GN are purposefully matched to these requirements, allowing integration with other techniques \cite{8472907} for providing low-latency services. In the context of 6G, GN needs to evolve towards encompassing involving sensory, positioning, and interaction/actuation data from the physical world, which should be seamlessly fused with the digital world.

The foreseen digital-physical world concept of networking infrastructure is illustrated in Fig.~\ref{network}, where sensing signals from the physical world are transmitted through the network to a central unit. Upon receiving signals, the central unit processes them and returns control signals to the virtual world. An actuation signal is then transmitted from the virtual to the physical world by the network. Among the various media that facilitate the interaction between the physical and digital worlds, such as wireless, wired, and touch interfaces, this paper focuses on the evolution of GN techniques to enhance the capabilities of 6G wireless networks, which are yet to be fully defined and standardized.

\section{Game Networking and Its Integration \\ within 6G}
This section describes fundamental aspects of the GN framework, delineating its architectural elements that possess the potential for evolution towards the forthcoming 6G network paradigm. Furthermore, we scrutinize two design considerations of next-generation networks that could greatly benefit from incorporating GN technology.
\subsection{Game Networking Overview} 
GN refers to the efficient transmission of data packets for exchanging updates between the users (clients) and a gaming server. The server acts as a centralized authority that is responsible for enhancing the user's experience and preventing tampering with the virtual environment \cite{smed2017algorithms}. This process is crucial for multi-user games and involves four key elements: state synchronization, entity interpolation, input prediction, and lag compensation. In GN, the system's state, i.e., the position and properties of entities in the virtual world, including the players' characters and other objects and agents, evolves according to the players' inputs and the internal logic of the game. The \emph{state synchronization}  optimizes the communication between servers and clients by prioritizing packets so that the most important information is delivered in a timely manner. \emph{Entity interpolation} attempts to minimize lag and jitter in the user experience by smoothing transformations and bringing them closer to the original movements. These phenomena often result from the less frequent transmission of updates or packet losses. This is supplemented by \emph{lag compensation} and \emph{input prediction} to create a smoother user  movement by anticipating the server's response to the input before it is received. 

GN operates in the network and transport layers, allowing for the scheduling and routing of data packets to establish paths for network communication. Academic literature and commercial products have offered various network architectural solutions to aid virtual applications with distributed users \cite{liu2022survey}. These solutions fall into three categories: Centralized, Distributed (P2P), or Hybrid, depending on the location of the state storage.  These architectural approaches effectively address the challenges of low latency and the high demands of real-time network traffic, demonstrating high potential for flexible evolution and integration into 6G wireless networks. This capability is particularly crucial in physical-virtual applications, where the transmission of large volumes of low-latency data is essential. Therefore, the improvements and applications of GN create blueprints for the progression of 6G networks towards supporting DT, VR/AR, and the Metaverse~\cite{9806418}.

\subsection{Essential Architecture Elements}
\subsubsection{Synchronization}
\begin{figure}[t!]
\centering
\includegraphics[width=0.49\textwidth]{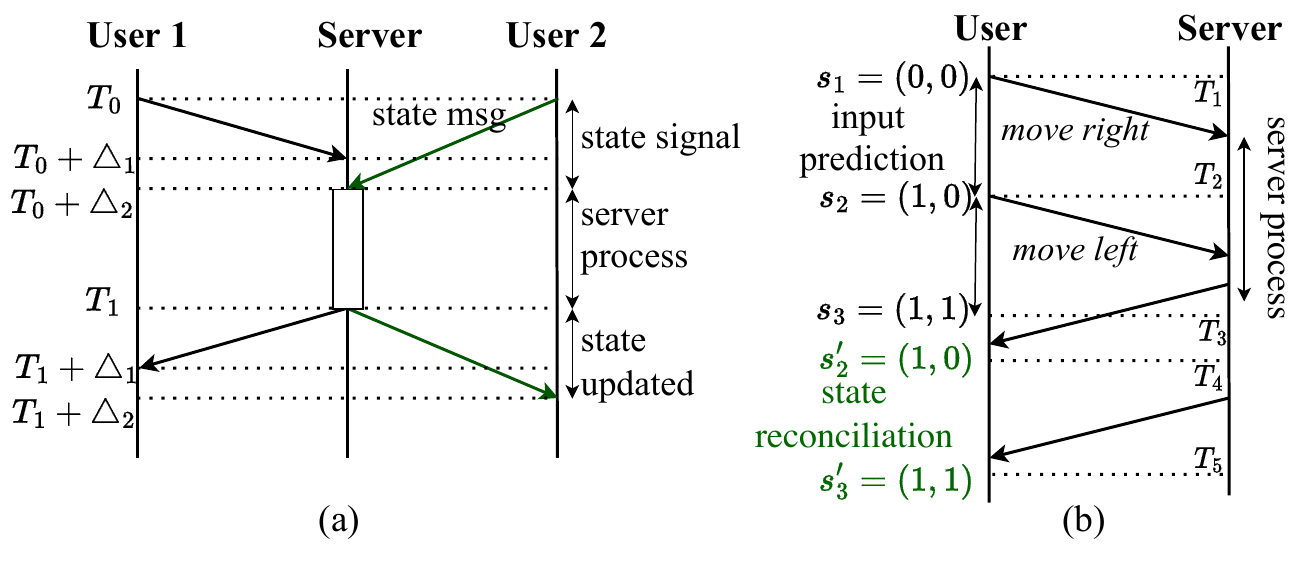}
\caption{Networking system with $(a)$ synchronization process of one server and 2 users; $(b)$ input prediction process where user predicts the output before receiving delayed true state from the server.}
\label{fig_sync_input}
\end{figure}
Tight GN-synchronization allows users to interact with each other in real-time seamlessly. The fulfilment of the user's QoE in 6G use cases entails synchronization of the physical and virtual realms within, while any mismatch in the user's recent state can result in a delayed response. Any alteration within the physical or virtual world must be reflected in the other  to be perceived in a seamless, ``natural'' way. 
To evaluate the synchronization between GN components, operations must be ordered and executed in the appropriate time frame. Real-time communication requires a reduction of out-of-order operation requests to nearly zero. If no synchronization is implemented, the latency and the jitter will most likely be high, making it essential for ensuring consistency and QoS in Virtual-like environments. 

Fig.~\ref{fig_sync_input}(a) illustrates a simple GN architecture with two users whose operations are coordinated and manipulated by the server. Here, two user states are synchronized with the server, with $\triangle_1$ and $\triangle_2$ being the latency of user 1 and user 2, respectively. If the processing only happens after the server gets all users' states ($T_0+\triangle_2$), the state updates of all users will depend on the user with the highest latency profile, which can lead to network inactivity for a longer time period, eventually violating the timing requirements. The complexity increases with multiple users and devices, each of which may access the same server/virtual world with a different latency.

State synchronization techniques in GN can be utilized to estimate and synchronize the physical-digital relationship, particularly when numerous sensors (e.g., accelerometers, cameras, magnetic, optical, and acoustic sensors) must be synchronized to instantiate a digital world.
The Precision Time Protocol (PTP) has been standardized, utilizing a ``grandmaster'' clock, such as a Global Navigation Satellite System, to distribute accurate time to leaf clocks~\cite{IEEE1588}, thereby supporting precise time synchronization. The heterogeneous synchronization requirements range from $1-10$~ns for communication, $100$~ps to $10$~ns for localization, and $100$~ps for sensing~\cite{wymeersch20226g}.  In industrial IoT systems, traditional methods like precision and flooding protocols require regular clock calibration. However, as the network scale expands, synchronization performance declines, impacting clock accuracy and increasing resource consumption. This underscores the need for new coordination methods, with GN-synchronization being a potential solution.

\subsubsection{Entity interpolation}
As a general rule, almost every multi-user game employs entity interpolation within its networking to minimize lags and jitters due to less frequent updates or packet drops. The interpolation process smooths out the transformations and increases their similarity to the original movements. This technique is not designed to anticipate future positions but rather to use only the actual entity data from the server, so other entities appear slightly delayed. 

Consider a scenario where several users send information simultaneously and at a fast pace (videos, commands, or movements). The CPU and bandwidth would be overburdened by updating the virtual world every time inputs are received from each user and then broadcasting the virtual state. In an ideal scenario, the user inputs are queued as they are received, without any processing, while the digital world is updated periodically at a low frequency. Entity interpolation is then purposefully employed to ensure that the motion appears continuous to the user. In VR/AR and DT systems, which are space- and time-sensitive and impact reality beyond entertainment, it is essential to conduct a more in-depth study of GN-based entity interpolation techniques to meet the reliability requirements. In such scenarios, advanced learning tools such as Generative AI (GAI) might bring a significant impact, which has to be investigated.

\subsubsection{Input prediction}
In online video games, predictions of inputs assist in forecasting and displaying player movement before the server responds to user commands. By combining the user's current state with the user's inputs, the next predicted state can be drawn. A user's state is adjusted once an update is received from the server, so the old data is deleted. 
This procedure is reiterated following each update to maintain a consistent and uninterrupted gameplay experience. Therefore, GN-related input prediction could be integrated into 6G infrastructure to ensure continuity and seamless end-user interactions, particularly in real-time use cases. 
In Fig.~\ref{fig_sync_input}(b), the user predicts states $s_2$ at $T_2$ and $s_3$ at $T_3$ using information from the current state and action commands. These states are then reconciled and corrected based on actuation signals from the server at $T_4$ and $T_5$. Input prediction ensures smooth, continuous movement of the display instead of sudden transitions, for example, from $s_1$ at $T_1$ to $s_2'$ at $T_4$.

\subsubsection{Delay/lag  compensation}
The client-server virtual architecture in computer games operates as follows: the server processes inputs from all users, including timestamps, updates the virtual world, and sends regular snapshots to all users. Users transmit their inputs, locally simulate their effects, receive updates on the virtual world, synchronize their predicted state to the authoritative state, and interpolate known past states for other entities simultaneously.  As a result, users see themselves in real-time but other entities in the past, leading to the use of GN-related delay compensation techniques to limit packet effects and improve the virtual-to-real scenarios.
The control system has shifted from traditional point-to-point to advanced network control, resulting in increased complexity with many communication nodes and network delays comparable to sampling intervals. This raises the demand to take appropriate measures and ensure the system's sustained performance and robustness.
Delay compensation techniques have been prevalent since the emergence of early time synchronization protocols such as the Network Time Protocol and PTP for wired networks and Reference Broadcast Synchronization and Flooding Time Synchronization Protocol for wireless sensor networks (WSNs).
These techniques have been further refined by incorporating methods based on round-trip delay measurement. 

These factors are individually or jointly considered in several 6G applications. However, adopting the GN concept, which enables cohesive integration of all aspects, particularly at the transport and network layers, allows disparate techniques and protocols to be integrated and coordinated efficiently.

\subsection{Design Considerations}
This part discusses two designs of the next-generation networks: \emph{real-time interaction (RTI)} and \emph{remote immersive interaction (RII)}. RTI refers to communication that occurs in the present moment with minimal delay, enabling participants to respond to one another in a near-instantaneous fashion. 
RII, on the other hand, concentrates on the reliability of the system at the end-user level, thereby guaranteeing a high-quality immersive interactions. This highlights the equivalence with the problems addressed by GN, leading to a natural application of GN principles to the networking that supports the fusion of the physical and digital realms.

\subsubsection{Real-time Interaction} 
RTI ensures that the players' actions and the game's responses are synchronized, providing an immersive and engaging gaming experience. 
Ensuring low-latency communication between players and the game server is crucial for enabling the game to respond swiftly to player actions, thereby facilitating a seamless user experience. With the emergence of cyber-physical systems, real-time status updates in networking are important and ubiquitous forms of communication. These updates include states, scores, traffic, sensing, and security system reports from remote systems integrated into virtual spaces. For online gaming, ``real-time'' is only an impression reflected through user dissatisfaction; for example, it is set to approximately $100$~ms for first-person shooter (FPS) games~\cite{claypool2006latency}. 
For motion-to-photon latency to be imperceptible, VR/AR developers and industry experts generally concur that the round-trip latency of an application should not exceed $20$~ms~\cite{mangiante2017vr}. As the distance increases, networks tend to experience an observable increase in latency, which can be as high as $100$~ms or even more. According to Michael Abrash (Oculus VR Chief Scientist), ``\emph{more than $20$~ms is too much for VR and especially AR, but research indicates that $15$~ms might be the threshold or even $7$~ms,}'' which calls for hardware improvements to bring display latency down\footnote{\url{https://www.gamesindustry.biz/valves-michael-abrash-latency-is-getting-in-the-way-of-vr/}. Accessed: Jun. 25, 2024}.

These architectures have common features: the source, which generates time-stamped status update messages, transmits these messages to one or more receivers through the communication system. However, the system's limitations dictate that the delivery of a status message requires a non-zero and often random time in the system.  
Changes in the digital object that induce changes in the physical object can be even more complicated. For instance, an important aspect of bringing DTs into manufacturing activities is the age of delivered up-to-date information and the required balance between the rate of updates against congestion. The proficient utilization of entity interpolation and delay/lag compensation techniques profoundly influences the reduction of the packet transmission load, thereby facilitating the preservation of transmission and reception quality. This serves the dual purpose of preventing network congestion and ensuring the fulfillment of stringent real-time communication demands. In such cases, the system can extrapolate information from sensor data and synchronize it with the actual physical environment, thus enhancing its adaptability and performance.

\subsubsection{Remote Immersive Interaction} 

In RII, a plurality of sensory information, such as visual, auditory, or haptic, is conveyed over long distances. The objective is to meet the reliability requirements and guarantee users' experience to enable a high-quality immersive interaction, regardless of real-time requirements. Within the domain of RII, the GN framework can be expanded to encompass considerations of reliability, and packet delivery jitter.

The accuracy of the reconstructed information should be maintained at a very high level. There is a possibility that packets may be received out of order, compromised, lost, or received with a significant delay. It is a common practice for certain network protocols (i.e., TCP), or error concealment and correction technologies  to implement mechanisms to ensure that all packets are conveyed in the correct sequence and within the expected timeframe. Achieving the high reliability required in 6G is associated with costs, e.g., packet retransmissions and feedback, consume additional bandwidth, which, along with feedback processing, further increases the overall system latency. 
Another major challenge involves scaling the size of a virtual environment while maintaining distributed consistency, such as sharing a sense of consistent spatial surroundings across a massive number of concurrent users.  To achieve this, each user must maintain a copy of the relevant virtual state on their hardware. When a user performs an action, the virtual states of other users, impacted by this action, are updated accordingly. As the number of online users and objects, as well as the complexity of these objects, increases, the volume of exchanged data grows exponentially. This further tightens the margin of error: a higher volume of data, with tighter timing and jitter requirements, makes end-to-end network optimization crucial.

On the other hand, the management of the delicate trade-off between latency, reliability, jiter, and large-scale distributed consistency has been effectively addressed within the realm of GN in the online game industry, accommodating diverse network conditions. This offers valuable insights to borrow the concepts of GN combined with advanced methods of processing, reliability to harness and extrapolate sensing signals. Specifically, a deterministic chain decision process is enforced by taking smart decisions within tight latency constraints and respecting the bounds, making it possible to introduce intelligence to enable new services. An example application is a remotely controlled high-precision manufacturing system, requiring a maximum jitter in the order of a microsecond \cite{8472907}. The proactive utilization of input prediction strategies serves the dual purpose of reducing foreseeable data redundancies within the network, thereby conserving transmission bandwidth. 
Furthermore, the implementation of delay and lag compensation mechanisms is instrumental in enhancing the overall fluidity and visibility of deterministic real-time monitoring applications.

\section{Case Studies}
In this section, we evaluate the integrated GN and communication network through two scenarios to illustrate examples of RII and RTI.  The first involves the transmission of real-time video through an ORAN network, facilitated by a GN that relies on GAI and is capable to adapt to the network conditions. The second one presents a trusted wireless DT actualizing the realization of RTI potential on Network Control Systems (NCS). 

\subsection{Demonstration of Entity Interpolation using Generative AI for Video}
The setup emulates the adaptability of the network to facilitate real-time video transmission through the ORAN network. Using GAI to represent the entity interpolation technique, the advanced communication system is able to preserve video quality while reducing communication resource usage. Specifically, we conduct experiments through service-based ORAN architecture, allowing a variety of vendors to play a part in the realization of the construction of a RAN. With the integration of mobile edge computing (MEC) this can be further enhanced by optimizing end-to-end performance of users via trading of communication and computational resources. ORAN itself creates a software-defined RAN, in which both monitoring and control of the RAN can be done on a near-real time level through the near real-time RAN Intelligent Controller (RIC). The RIC exposes service models (SMs), which can be leveraged by xApps for control and monitoring of KPIs.

\begin{figure}
    \centering
    \includegraphics[width=0.48\textwidth]{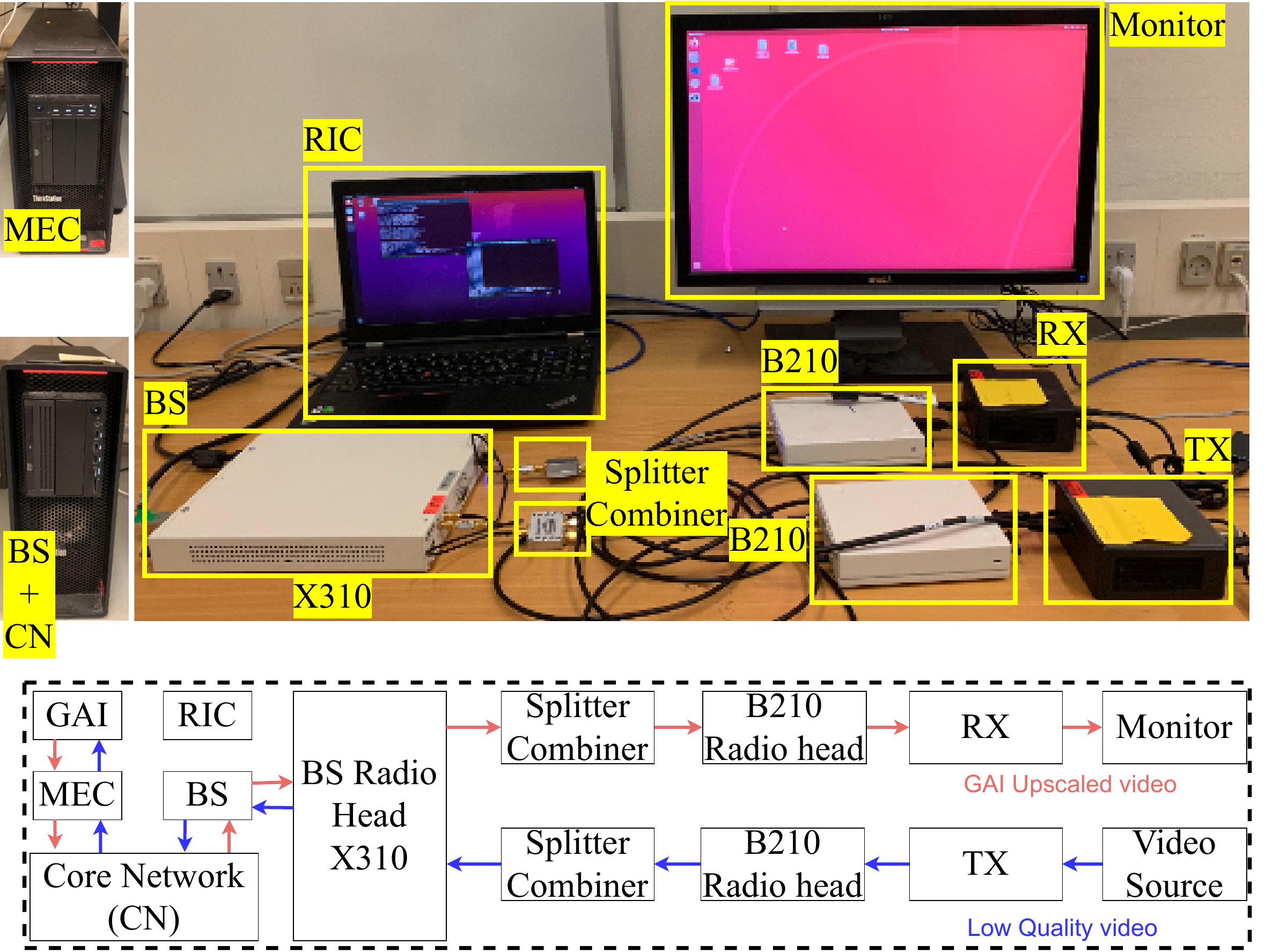}
     % \vspace*{-0.2cm}
    \caption{Experiment setup.}
    \label{fig:lab-setup}
    % \vspace*{-0.5cm}
\end{figure}

We evaluate this experimentally in a laboratory, see Fig.~\ref{fig:lab-setup}. A RAN is orchestrated via the OpenAirInterface project for the base station, srsRAN for the UEs and open5gs for the core network. The RIC is implemented as the FlexRIC project. Due to the limitation of the FlexRIC project this is implemented on 4G, but given the ORAN framework this can be extended to 5G and future cellular generations. We utilize a computer equipped with augmented processing power and dedicated graphic cards to facilitate the execution of the GAI model. Within our experimental setup, TX's traffic is directed to the MEC. The MEC, discerning whether the received content has undergone a reduction in picture quality, applies its GAI accordingly. In cases where the picture quality has been compromised (caused by channel capacity), the SRGAN module \cite{ledig2017photo} are employed to enhance the video quality before forwarding to the designated recipient to maintain the high quality real-time experience to users.  We conducted experiments employing the MOB dataset \cite{Ahmed_2023}, specifically scrutinizing a random selection of 50 videos from the dataset.

\begin{figure}
    \centering
    % \vspace*{-0.1cm}
    \includegraphics[width=0.49\textwidth]{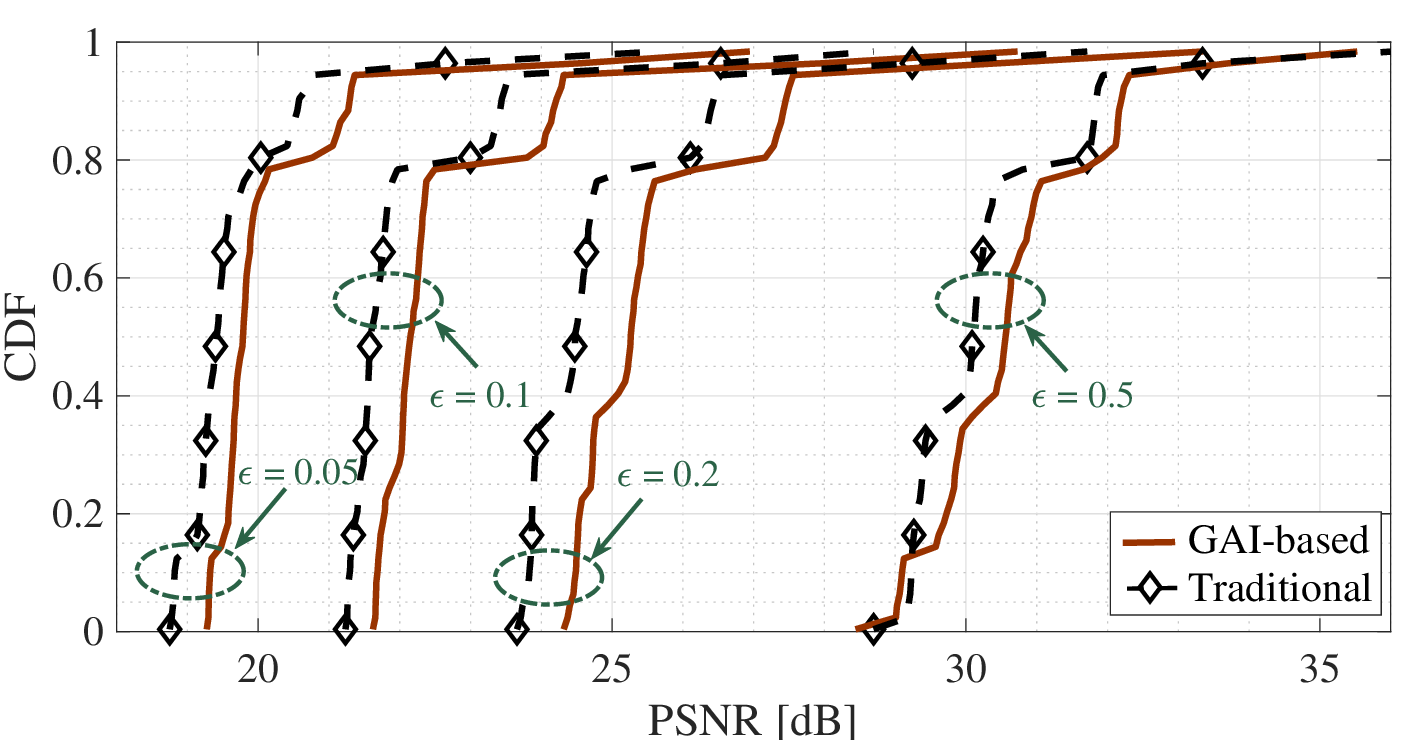}
    % \vspace*{-0.1cm}
    \caption{CDF of PSNR under different resolution factor ($\epsilon$). }
    \label{fig:oran-experiment}
    % \vspace*{-0.5cm}
\end{figure}

Fig.~\ref{fig:oran-experiment} illustrates the CDF of Peak signal-to-noise ratio (PSNR) between received and transmitted frames by deploying our strategy (\textit{GAI-based}) versus the \textit{Traditional} scheme (normal transmit and receive frames) under different resolution factor ($\epsilon$). Specifically, with $\epsilon = 0.2$, there are 80\% frames of \textit{Traditional} with a PSNR below $35$~dB. Thanks to the advanced GAI-based, the proposed architecture provides around 40\% of the frames having a PSNR below $35$~dB, which is two times better than the baseline. This result confirms that integrating GAI into the ORAN network can bring high efficiency in terms of PSNR. Furthermore, under circumstances characterized by the high cost of communication resources, this established configuration proves advantageous in mitigating the volume of data transmitted across the network. Such optimization facilitates resource conservation for other applications. This case study illustrates how the ORAN framework can utilize new techniques in its service-based architecture to create a more flexible network architecture, thereby better adapting to real-time applications of 6G wireless networks.

\subsection{Trusted Wireless Digital Twin for monitoring, prediction and controlling} 
\begin{figure}[t]
    \centering
    \includegraphics[trim=0cm 0.0cm 0.cm 0cm, clip=true, width=0.49\textwidth]{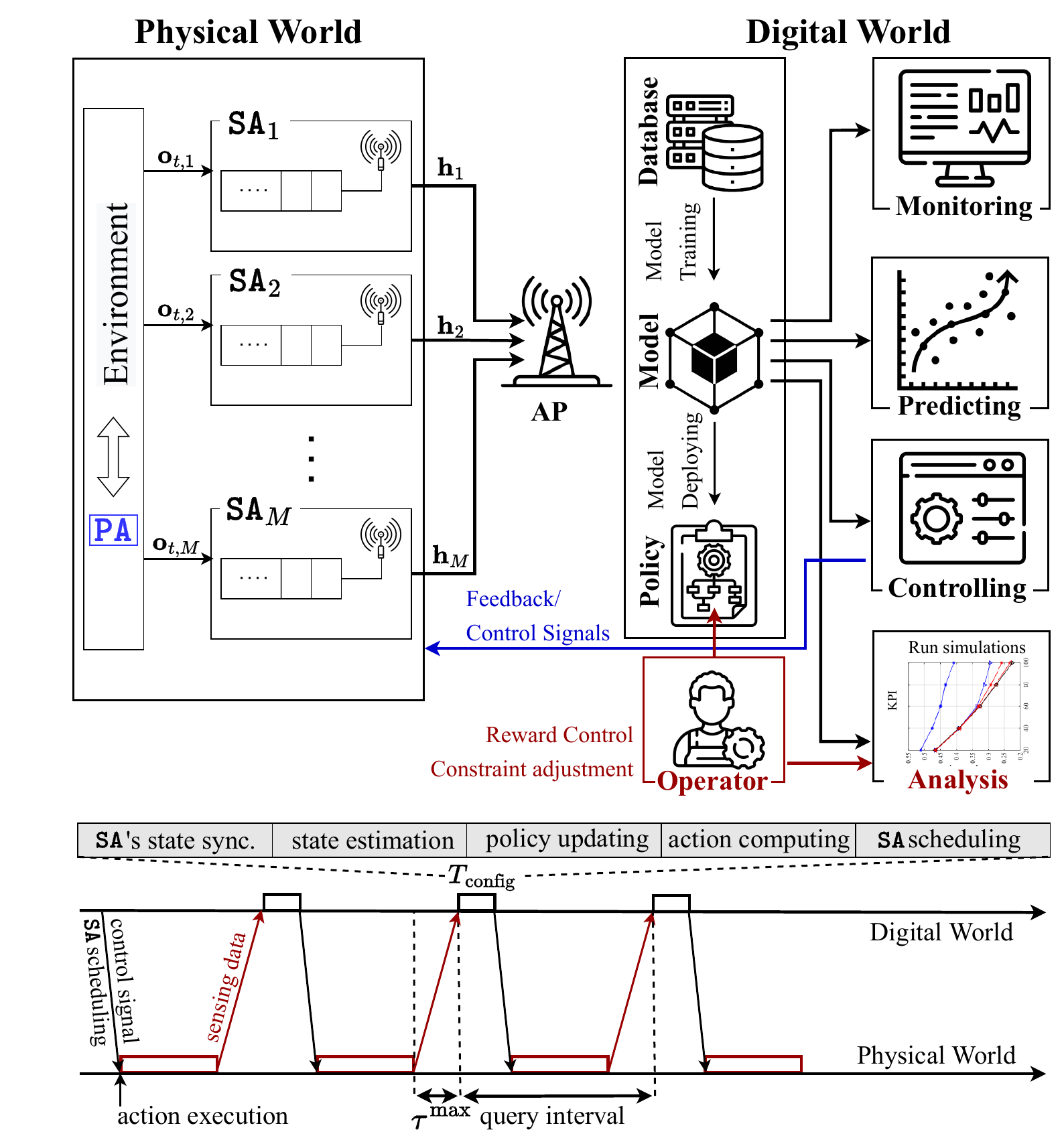} %Communication_diagram.pdf
    \caption{\phuc{The considered architecture depicting fusion of physical and digital worlds with communication diagram.}}
    \label{fig_system}
    % \vspace*{-0.5cm}
\end{figure}

This setup  consists of sensors in a NCS that are used to build a DT model of the system dynamics. The focus is on control, scheduling, and resource allocation for sensory observation to ensure timely delivery to the DT model deployed in the cloud. Low latency and communication timeliness are instrumental in ensuring that the DT model can accurately predict system states and optimize control policy. 
Particular, the DT architecture and corresponded communication diagram are illustrated in Fig.~\ref{fig_system}, including a single primary agent ($\PA$) and a set of sensing agents ($\SA$s). These $\SA$s are responsible for observing the environment and  communicate with the \emph{access point} (AP) through a  wireless channel, which facilitates the construction of the DT model for the $\PA$ and operates in Frequency Division Duplexing (FDD) mode. 
At the beginning of each \emph{query interval} (QI), the DT model is required to update the active state of $\SA$s, after which it estimates the full system state, updates policy, computes optimal control signal, schedules at most $10$ $\SA$s and applies a fusion algorithm at the AP. $\SA$ selection for observing its sensing signals is done through the revised Extended Kalman Filter, guided by the prediction acuracy requirements of DT. 
This is inspired by GN-based entity interpolation, performed as part of the estimation process, needs to be matched with input prediction to provide an up-to-date estimate. 
Then, we propose an uncertainty control actor-critic reinforcement learning (RL) with Proximal Policy Optimization to devise optimal controls for the physical world and effectively managing the state estimation errors \cite{NIPS1999_6449f44a}.  We note that, unlike the state's certainty requirements at the DT, which are often set by the Operator and fixed, the RL agent for controlling the $\PA$ receives input as an estimated state with varying certainty, adjusted based on the current $\PA$'s state, and output the control signal and required observation's accuracy for the next time step.
Once a control command is generated, the controller promptly transmits it through a downlink channel to the physical world. The application output for actuator control retrieves the most recently stored command values from memory and applies them to drive the system dynamics.  Upon completion of the task, the scheduled $\SA$s sample the state and subsequently transmit it to the DT via the AP within a latency threshold lower than $\tau^\mathrm{max} = 5$ms. 

\begin{figure}
	\centering
	\includegraphics[trim=0.0cm 0.0cm 0.cm 0cm, clip=true, width=0.49\textwidth]{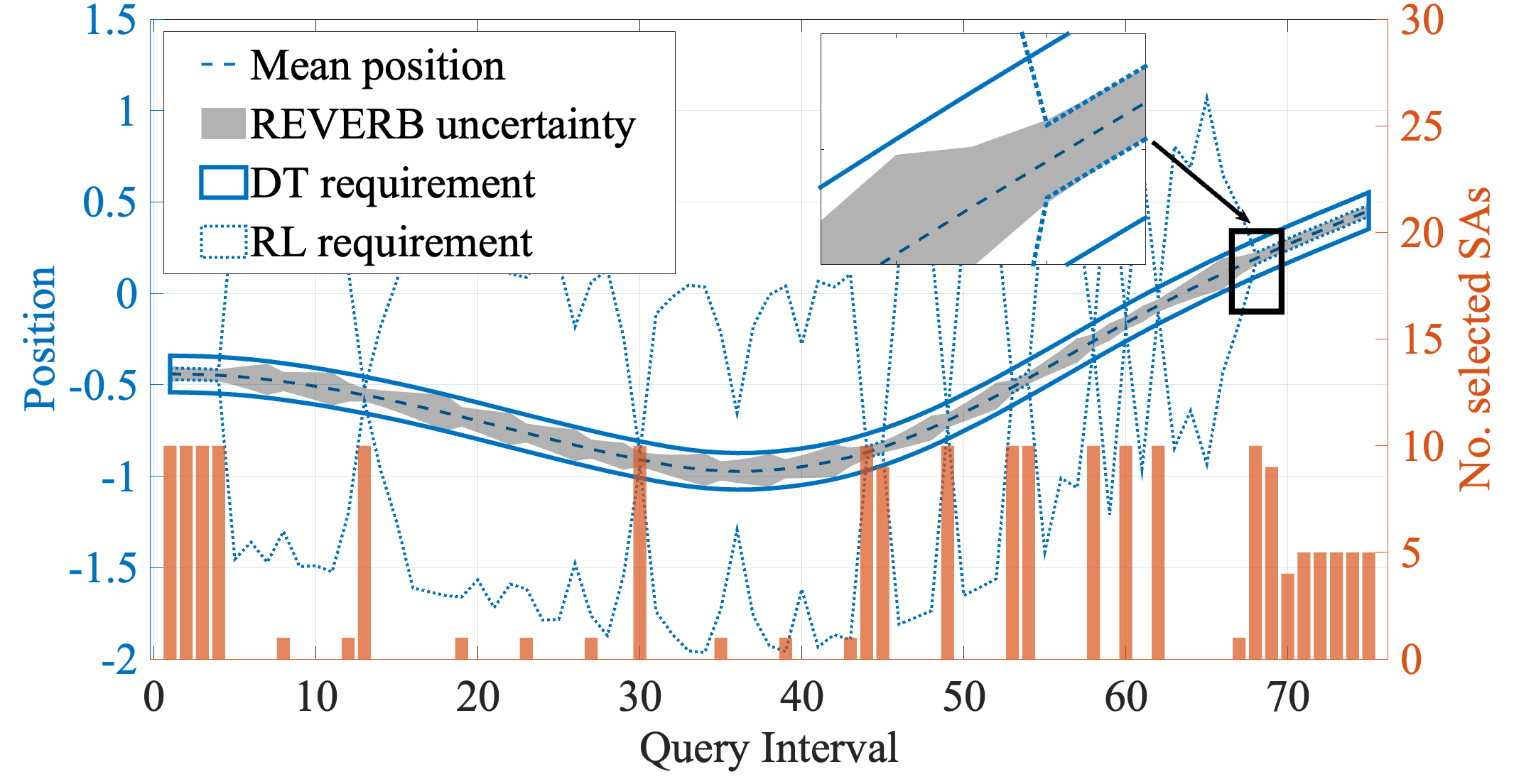}
	% \vspace*{-0.4cm}
	\caption{Uncertainty evolution and number of selected $\SA$s vs QI.}
	\label{DT_resutls}
	% \vspace*{-0.5cm}
\end{figure}

Given the limitations of the network, it is impossible for the DT to get data from all $\SA$s at any time to provide a reliable replica of physical world. 
The considered DT framework, denoted as REVERB (REinforcement learning and Variational Extended Kalman filter with Robust Belief), is employed to address the ContinuousMountainCar-v0 environment from the OpenAI Gym\cite{moore1990efficient}. Our results are presented in Fig.~\ref{DT_resutls} explaining the uncertainty evolution and the strategic selection of $\SA$s based on their contributions to the DT's performance. It is noteworthy that the management of REVERB's uncertainty is subject to the control of both DT and RL requirements. 
Hence, REVERB only requests more observations when the DT threshold is surpassed or when the RL agent necessitates high precision, typically when the agent is nearing its goal and precise force control is imperative.  
We emphasize that by taking advantage of input prediction and entity interpolation inspired by GN, DT maintains a reliability of estimation while allowing for the conservation of communication resources, which can be allocated for other services.

\section{Research Opportunities and Challenges}
This section explores the opportunities and challenges of adapting GN to 6G-enabled physical-digital framework. We present scenarios employing GN techniques to efficiently manage resources during interactions, leveraging system-level knowledge to bridge the physical and digital worlds.

\subsection{ML-Edge}
Network infrastructure must improve decision-making speed and reliability at the edge to meet the demands of real-time mission-critical applications. Currently, MEC systems are congested in the transport network and do not provide users with an efficient deterministic service. To address this, intelligence needs to be pushed to the edge with tight coordination between network and computation resources. 
Typical requirements include an uplink bandwidth exceeding $50$~Mb/s, a maximum delay of $1$~ms, and reliability of greater than $99.9999\%$~\cite{rfc8578}. 
Ensuring sufficient QoS for machine learning (ML) applications using distributed computing and networking technologies is challenging. \phuc{This raises the demand for integrating input prediction from GN to preemptively anticipate the state of the network and system resources, allowing for quicker decision-making and reducing perceived latency.  For instance, by predicting the next set of actions or data inputs, the system can prepare resources in advance, ensuring smoother operation and reducing delays. Additionally, lag compensation techniques can be applied to maintain consistency and reliability in edge computations, even when network conditions fluctuate. This helps ensure that tasks executed in use cases such as data aggregation, calculation, and structure metadata creation, meet throughput and latency limits to guarantee AoI and reliability constraints.}
\subsection{Security Problems} 
The privacy and security of data in a virtual environment are major concerns, even though organizations have proceeded to innovate their security systems since Metaverse networks are shared virtual environments where numerous diverse users build. 
From a network perspective, security issues frequently occur in communication processes. Since the transmission contains information regarding the behaviour and identity of the user, malicious attackers can deduce sensitive information based on unwittingly leaked information. 
\phuc{Accordingly, input prediction could be utilized to detect abnormal patterns in data transmission, which may indicate security threats, allowing for preemptive action to be taken. Besides, DT can be employed to test and reinforce security mechanisms by creating virtual environments where potential security breaches are simulated and analyzed. This allows for the identification and mitigation of vulnerabilities before they can be exploited in real scenarios.}
\subsection{Distributed Consistency}
Presenting users with a unified shared virtual environment is complicated by potential discrepancies in the timing of updates communicated to different nodes. Furthermore, if participants act based on different perceptions of the environment, a mechanism must be established to determine the accurate version. Inconsistency negatively impacts user immersion, diminishing overall experience and potentially reducing the inclination to continue using the system. 
\phuc{Achieving true consistency within a distributed environment is a complex task, requiring the consideration of subtle failure modes related to timing and ordering. Here, entity interpolation and input prediction  can be leveraged to maintain consistent states across distributed nodes, even under varying network conditions. By interpolating the state of virtual objects based on predicted inputs, it is possible to ensure that all users have a coherent view of the virtual environment, reducing the impact of latency and network jitter. Moreover, lag compensation helps to align the timing of updates across different network segments, ensuring that all users experience the environment in a synchronized manner.}

\subsection{Information Timeliness and Semantic/Pragmatic Communications}
The scalability of a network is restricted by the communication capability of its deployment, as coordinating multiple parts demands communication. Ideally, all nodes communicate simultaneously without congestion, which calls for a network architecture to manage the traffic. 
\phuc{Transmitting the state of every user to all participants in a multi-player virtual environment is inefficient, presenting the challenge of selectively transmitting only essential information. Synchronization and lag compensation have the potential to help optimizing the timing of data packet transmission, ensuring efficient and timely updates across the network.}

The paradigms of \emph{semantic and pragmatic (effective) communication} can be valuable towards identifying such important information. Semantic techniques interpret user-specific characteristics, such as profiles, preferences, and actions, rather than the entirety generated messages. These interpretations enable proper orchestration of networking interactions, thus enhancing the overall experience.

%%%%%%%%%%%%%%%%%%%%%%%%%%%%%%%%%%%%%%%%%%%%%%%
\section{Conclusion}
The article provided GN principles, highlighted key design considerations, and opened research challenges in realizing them with the 6G network. We show how networking design plays a crucial role in facilitating a paradigm shift towards real-time and remote immersive interactions. Two case studies are presented, showcasing the potential of GN techniques to alleviate stress on communication systems while maintaining the necessary quality during user interactions.

 % \setstretch{0.9}
\bibliographystyle{IEEEtran}
%\balance
\bibliography{Journal}

\section*{Biographies}
% \vspace{-0.5cm}
\begin{IEEEbiographynophoto}{Van-Phuc Bui}
received the B.Sc. degree (Hons.) in Electronic systems from the Ho Chi Minh University of Technology, Vietnam, in 2018 and the M.Sc. degree in telecommunication from Soongsil University, South Korea, in 2020. He is currently pursuing a Ph.D. at Aalborg University, Denmark. His research interests include wireless communications, semantic communications and networking.
\end{IEEEbiographynophoto}
% \vspace{-0.7cm}
\begin{IEEEbiographynophoto}{Shashi Raj Pandey} is currently an assistant professor at the Connectivity Section, Aalborg University, Denmark. He received his Ph.D. in computer science and engineering from Kyung Hee University, South Korea, in 2021. His research interests include network economics, wireless communications and networking, distributed machine learning, and semantic communications.
\end{IEEEbiographynophoto}
% \vspace{-0.7cm}
\begin{IEEEbiographynophoto}{Andreas Casparsen}
received the B.sc. degree in Electronic systems, from Aalborg University, Denmark in 2019. He got his degree in Communication engineering from Aalborg University, Denmark, in 2021. Currently, he is pursuing a Ph.D. at Aalborg University, Denmark. His research interest includes ORAN, latency, and experimental research.
\end{IEEEbiographynophoto}
% \vspace{-0.7cm}
\begin{IEEEbiographynophoto}{Federico Chiariotti}
is currently an assistant professor at the University of Padova, Italy, where he also received his PhD in information engineering in 2019. From 2020 to 2022, he worked at Aalborg University, Denmark. He has authored over 90 peer-reviewed papers on latency optimization, semantic and goal-oriented communication, and machine learning in networking. 
\end{IEEEbiographynophoto}
% \vspace{-0.7cm}
\begin{IEEEbiographynophoto}{Petar Popovski}
is a Professor at Aalborg University, where he heads the section on Connectivity and a Visiting Excellence Chair at the University of Bremen. He is the Editor-in-Chief of IEEE JOURNAL ON SELECTED AREAS IN COMMUNICATIONS. His research interests are in communication theory and wireless connectivity.
\end{IEEEbiographynophoto}

\end{document}